\documentclass{article}
\usepackage{ijcai26}

\usepackage{times}
\usepackage{soul}
\usepackage{url}
\usepackage[hidelinks]{hyperref}
\usepackage[utf8]{inputenc}
\usepackage[small]{caption}
\usepackage{graphicx}
\usepackage{amsmath,amssymb}
\usepackage{amsthm}
\usepackage{booktabs}
\usepackage{algorithm}
\usepackage{algorithmic}
\usepackage[switch]{lineno}
\usepackage{enumitem}
\usepackage{tikz}
\usetikzlibrary{calc}
\usetikzlibrary{positioning}
\usetikzlibrary{shapes.geometric}
\usetikzlibrary{arrows.meta}

\newtheorem{example}{Example}

\newtheorem{remark}{Remark}

\def \S{\mathcal{S}}

\def \M{\mathcal{M}}

\def \V{\mathcal{V}}
\def \C{\Rn(\U)}
\def \U{\mathcal{U}}
\def \Rn{\mathcal{R}}

\def \F{\mathcal{F}}

\def\c{\vec{u}\hspace{.1ex}}

\def\O{\mathcal{O}}
\def\g{\mathbf{g}}

\def \m{M}

\def \<{\langle}
\def \>{\rangle}

\newcommand{\hide}[1]{}

\def \util{\small\textsc{Util}}
\def \val{\small\textsc{Val}}

\DeclareMathOperator*{\argmax}{arg\,max}
\newcommand{\emoji}[2]{
    {\includegraphics[width=#2ex]{#1.png}}
}

\newcommand{\pray}{\emoji{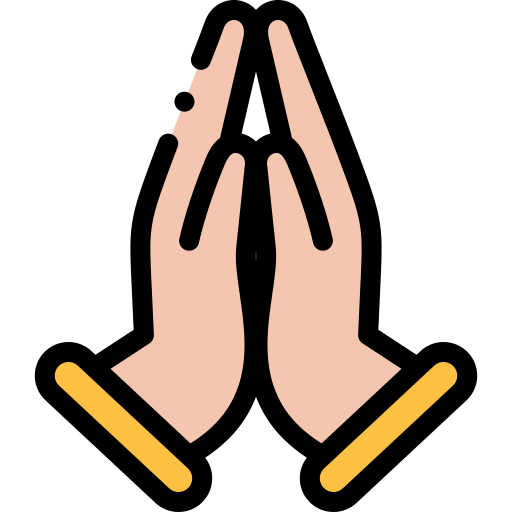}{2}}
\newcommand{\sun}{\emoji{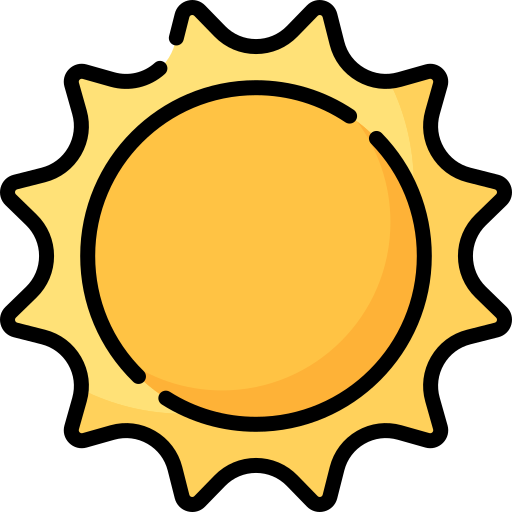}{1.8}}

\newcommand{\snake}{\emoji{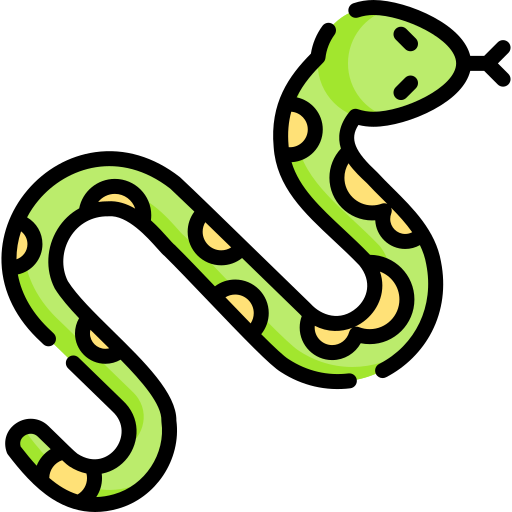}{2}}
\newcommand{\bounty}{\emoji{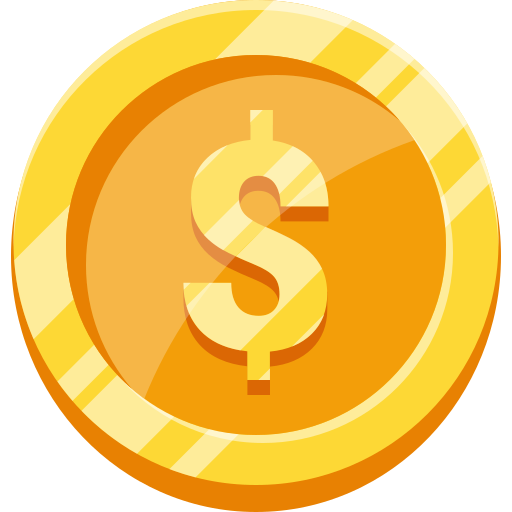}{1.8}}
\newcommand{\eggs}{\emoji{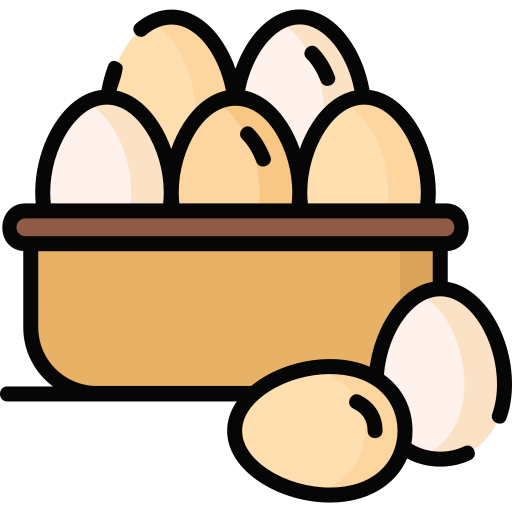}{1.8}}
\newcommand{\hunt}{\emoji{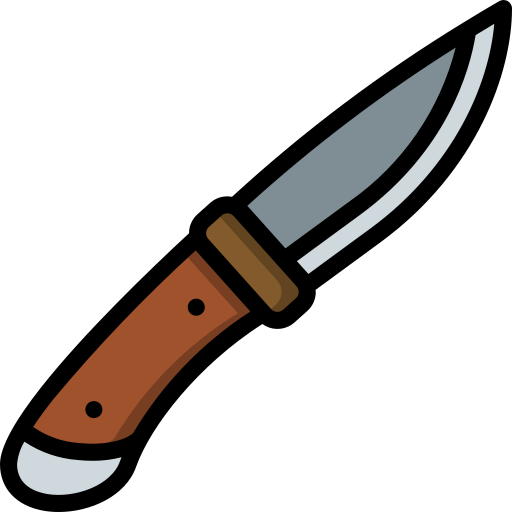}{2}}

\newcommand{\eggsbw}{\emoji{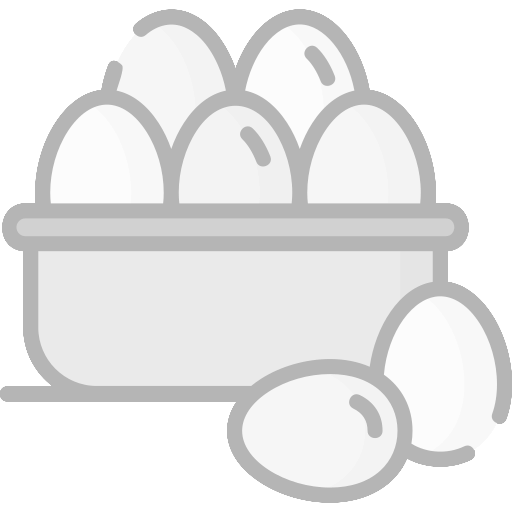}{1.8}}
\newcommand{\commentout}[1]{}

\title{Unintended Consequences: Updating Causal Models}

\author{
Joseph Y. Halpern$^1$
\and
Evan Piermont$^2$
\and
Marie-Louise Vier\o$^{3}$\\
\affiliations
$^1$Computer Science Department, Cornell University, Ithaca, USA.\\
$^2$Department of Economics, Royal Holloway, University of London, UK.\\
$^3$Department of Economics and Business Economics, Aarhus University, Denmark.\\
\emails
halpern@cs.cornell.edu,
evan.piermont@rhul.ac.uk,
mlv@econ.au.dk
}

\begin{document}

\maketitle

\begin{abstract}
We examine how causal beliefs affect an agent's
  choices and how feedback on those choices leads to updated causal
  beliefs. Building on the structural-equations framework for modeling
  causality, we first examine the general problem of updating causal
  beliefs in the face of novel (and possibly inexplicable) data. We
  model an agent who is uncertain of the true causal model, and
  therefore entertains a probabilistic belief over the set of possible
  models. We then consider how causal beliefs influence choices by
  building a model of agency and utility on top of the usual
  structural-equations framework. Using these two components, we
    propose a notion of \emph{steady state}, where the feedback
  received from an agent's optimal action, given her current beliefs
  about the true causal model, can be rationalized by those beliefs.  
\end{abstract}

\section{Introduction}

Causal reasoning is difficult; agents routinely misunderstand,
misinterpret, or simply overlook the true causal relationships between
variables of interest. Examples of such erroneous judgments are
plentiful, ranging from the pedestrian (superstitiously carrying a
lucky charm) to the existential (misjudging humanity's effect on
global climate). Because an agent will decide which actions to take
based on her understanding of the \emph{effect} of her actions,
errors in causal reasoning are costly, leading to suboptimal or
inefficient choices. 
An agent may entertain different causal models, 
each reflecting a different understanding of the causal relationships, 
and hold probabilistic beliefs over these. 
In dynamic or repeated environments, an agent who entertains
incorrect causal judgements might observe evidence that contradicts
her initial understanding, requiring her to re-evaluate her beliefs. 
However, as we show, this re-evaluation may not result in correct
beliefs or stable behavior, even after an infinite number of interactions.

In this paper, we examine more generally how causal reasoning
affects an agent's choices and how feedback on those choices leads to
updated causal beliefs. In order to formalize causal reasoning, we
draw on the literature on structural causal models \`a la
\cite{pearl:2k}. 
We first examine the general problem of updating causal beliefs in the
face of novel (and possibly inexplicable) data. We model an agent who
is uncertain of the true causal model, and therefore entertains a
probabilistic belief over the set of possible models. To model the
agent's reaction to inexplicable evidence (that is, when she observes
outcomes that cannot be explained by any model in the support of her
beliefs), we endow the agent with a \emph{conditional probability
system (cps)} \cite{Popper34,Popper68,Finetti36}, encoding the agent's
behavior conditional on any 
(including potentially probability 0) observation.\footnote{We interpret the
beliefs conditional on probability 0 events as the analyst's model of
the agent's behavior, and not as beliefs consciously held by the
agent. In other words, the agent need not anticipate how she will
react to inexplicable evidence.}

We then consider how causal beliefs influence choices by building a model of agency and utility on top of the usual structural-equations framework. We model an agent who can take actions that consist of intervening on a set of variables and making partial observations about the outcome of the intervention. The agent derives utility as a function of the values of the variables, and thus her ability to intervene has a direct consequence on her utility. In dynamic environments, the agent also derives value from her observation regarding the effect of interventions, as such information forms the basis of belief updating, and therefore allows for better future decision making.

Using these two components, we propose a notion of
\emph{steady state}, where the feedback received from an agent's
optimal action, given her current beliefs about the true causal model,
can be rationalized by those beliefs. In such a case, the optimal
action does not induce the agent to revise her beliefs, and therefore,
the action remains optimal. To get a sense of the relation between an
agent's choice of action, the data that is generated, and the
re-evaluation of beliefs, consider the following two
examples. In both, the protagonist takes the incorrect action
(as judged from the modeler's perspective) on the basis of their
faulty causal reasoning; the difference lies in how the agents react
to the observed evidence: 

\begin{example}
\label{ex:ggIntro}
During
British East India
Company rule of colonial India, to combat the proliferation of
dangerous  serpents, the Governor came up with what appeared
an ingenious solution: pay a bounty to any man who brought in the
head of a dead cobra.  This ``cash for snakes” program worked for a
while, but was followed by a marked uptick in the number of cobras
plaguing the city. Indeed, entrepreneurial locals began breeding
snakes for the express purpose of collecting the bounty, but, owing to
lax security, many snakes escaped into the wild, dramatically growing
the population.  
\end{example}

The Governor \emph{learns} that his action was misguided, since the observation that arises subsequent to his choice of action cannot be rationalized by his initial understanding. Our second example, by contrast, shows that an agent  can remain in perpetual delusion, if the evidence generated by her action accords perfectly with her beliefs, despite the fact that those beliefs are incorrect.

\begin{example}
\label{ex:aitionsIntro}
Since time immemorial, members of the \emph{Aition} tribe have woken before sunrise to perform an elaborate ritual to their God \emph{Apotelesma}. They have never missed a single day, for they operate under the (fallacious) conviction that Apotelesma is vindictive and petty, and without their daily votive, she would not allow the sun to rise. Every day, in the daylight after their morning prayer, the Aitions conclude their model of the world must be correct as it accurately predicted the sunrise. 
\end{example}

\subsection{Related Literature}

Understanding the causal relationships between variables has been the focus of considerable work across many fields of study (see, for example,
\cite{angrist2009mostly,cunningham2021causal,HR20a,MW07,parascandola2001causation,plowright2008causal,Pearl09,pearl:2k,SpirtesSG},
noting that this list is very cursory). Despite the considerable attention paid to both modeling and uncovering causal relationships in general, there has been much less focus on modeling subjective causal reasoning as a component of decision making. Notable exceptions include 
\cite{schenone2020causality,ellis2021subjective,alexander2023subjective}, who all relate subjective notions of causality to decision making, although not within the structural-equations framework. 

Halpern and Piermont \shortcite{HP24a}
consider an agent who has a preference relation
over interventions and provide axioms characterizing ``causal
sophistication,'' that is, when the agent's preferences can be
rationalized by a utility function over variable values and a belief
over structural equation models. Their representation theorem can be seen as a
decision-theoretic foundation for the current framework, showing that the
agent's  utility function and (current) beliefs can be elicited from behavioral
data.

Psychologists have studied how humans actually learn causal relationships. 
In simple, controlled environments, there is ample evidence that
subjects' update their beliefs in a manner that approximates
Bayesianism, albeit under some computational constraints
\cite{griffiths2005structure,kushnir2007conditional,bonawitz2014win,coenen2019testing}. It
is less clear, however, how belief formation works in larger and less
controlled settings, where the space of potential causal models is too
large or unstructured to be considered all at once, or where the agent
is simply unaware of some relevant relationships. 
Bramley
et al.~\shortcite{bramley2017formalizing} argue that people
keep things simple, and
``maintain
only a single hypothesis about the global causal model, rather than a
distribution over all possibilities and update their hypothesis by
making local changes'' (where, by ``local changes'', they mean ``small
changes to the causal model under consideration'', such as adding
more parents to a node or changing the direction of causality in one
case).

Our paper lies within the more general literature on misspecified
models, where an agent persistently entertains the wrong model
governing data generation or the resolution of uncertainty.  This topic
has seen considerable recent interest within economics; see, for
example,
\cite{mailath2020learning,eliaz2020cheating,ellis2021subjective,cerreia2025making}.  
A consistent theme in these papers is that agents' beliefs will often
fail to converge (and in particular,
fail to
converge to the truth). We contribute to this literature by focusing
on two novel aspects: first,  
we model an agent's causal judgments, that is, her beliefs about the
effect of her 
actions. Second, we consider the possibility that the agent abandons
her initial beliefs in the face of inexplicable evidence.
Model-misspecification has also been studied in strategic
environments
\cite{fudenberg1993self,eyster2005cursed,esponda2016berk,zuazo2026misspecified}.  
While we do not consider strategic environments here, the extension of
our framework to this setting seems straightforward.

Our approach also bears similarity to the literature on belief
updating in the face of unawareness, e.g.,
\cite{karni2013reverse,piermont2021unforeseen,HP20}. In
particular, we can interpret an agent who systematically neglects the
causal influence of a particular variable as being unaware of the
variable (or at least unaware of its relevance). For example, we could
interpret the Governor in Example \ref{ex:ggIntro} as being
unaware of the possibility that the local population might breed
snakes.  

\section{Model}

We begin by describing the general framework of structural causal
models and how we model an agent's beliefs about them. At a high
level, we assume a dynamic environment, where, in each period, a causal
model stochastically determines the realization of all variables. The
agent does not know the true causal model, but entertains
probabilistic beliefs about it. Each period, the agent derives utility
directly from the realized values of variables, and in addition, can
take actions to intervene on variables
and (partially) observe the effect of her
intervention. Through this process she updates her beliefs. 

The
three main ingredients of our model, described in detail in the
remainder of this section are: (a) a universe of variables and
possible causal relations between them embodied by a set of possible
structural equations; (b) a conditional probability system on this
space of all structural equations that governs the agent's beliefs
(and eventually her behavior); (c) the actions that are available to
the agent and a notion of utility that directs the agent's choice of
action. 

\subsection{Structural-Equations Models}

We assume that the world is described by a set of variables. It is
useful to split them into two sets: the \emph{exogenous} variables,
whose values are determined by factors outside the model, and the
\emph{endogenous} variables, whose values are determined by the
exogenous variables and other endogenous variables.  We assume that
agents can intervene only on endogenous variables (but not necessarily
all of them).

Let $\U$ and $\V$ denote the (finite) collection of exogenous and endogenous variables, respectively, that describe the environment. For each $X \in \U \cup \V$, let $\Rn(X)$ denote the (finite) set of values $X$ can take. 
A \emph{context} is a vector $\c$ of values for all the exogenous
variables $\U$. Let $\C = \prod_{Y \in \U } \Rn(Y)$ consist of all
contexts.  

We use the standard vector notation to denote sets of variables or their values, that is, if $\vec{X} = (X_1, \ldots, X_n)$ and $\vec{Y} = (Y_1, \ldots, Y_m)$ are vectors, we write, in a standard extension of the usual containment relation, $\vec{X} \subseteq \U \cup \V$ to indicate that each 
$X_i \in \U \cup \V$, 
or, $\vec{X} \subseteq \vec{Y}$ to indicate that for each $i \leq n$, there is some $j \leq m$ such that $X_i = Y_j$. Similarly, we write $\vec{x} \in \Rn(\vec{X})$ if $\vec{x} \in \prod_{i \leq n} \Rn(X_i)$. As with the notation for random variables, uppercase letters refer to the names of variables and lowercase to their values. 
If $\vec{X} \subseteq \vec{Y}$ and $\vec{y} \in \Rn(\vec{Y})$, let
$\vec{y}|_{\vec{X}}$ denote the restriction of $\vec{y}$ to
$\vec{X}$. In particular, for a single variable $X$, $\vec{y}|_{X}$
denotes the $X^{th}$ component of $\vec{y}$.   


A \emph{causal model} is a pair $\m =  \<\S, \F\>$, where $\S =
(\U,\V,\Rn)$ is a \emph{signature} 
and $\F$ is a collection of
structural equations that determine the values of endogenous variables
based on the values of other variables.   Formally, $\F = \{F_X\}_{X
  \in \V}$, where $$F_X: \prod_{Y \in \U \cup (\V - \{X\})} \Rn(Y) \to
\Rn(X).$$  Given $\F$, we say that $X$ \emph{depends on} $Y$ if, there is
some setting of the variables in $\vec{Z} = (\U \cup \V) -\{X,Y\}$ and
values $y,y' \in \Rn(Y)$ such that $F_X(\vec{z},y) \ne
F_X(\vec{z},y')$; that is, changing the value of $Y$ changes the value
of $X$.  A causal model  is \emph{acyclic} if there are no cyclic
dependencies.  Following the literature, we restrict attention in this
paper to acyclic models.

We can represent qualitative features of a causal model by a
\emph{causal graph}, where the nodes are labeled by distinct
(exogenous or endogenous) variables, and there is an edge from $X$ to
$Y$ if $Y$ depends on $X$.  Note that the graph of an acyclic model is acyclic.
Let $\M$ denote the set of all acyclic models (note that $\M$ is finite).



  A \emph{setting} is a pair $(\m,\c)$ consisting of a model and a context.
  A setting $(\m,\c)$ uniquely
  determines the value of all variables. That is, there is a unique value
for each variable that simultaneously satisfies all the structural
equations in $\m$ and agrees with the context $\c$. We write $(\m,\c) \models
\vec{Y} = \vec{y}$ for the vector $\vec{y} \in \Rn(\vec{Y})$ if
$\vec{y}$ is the value of $\vec{Y}$ in $(\m,\c)$.

See \cite{Hal20} for more details
about the semantics of structural models. 

We denote the intervention on variable $X \in \V$ that sets its value
to $x \in \Rn(X)$ by $X \gets x$, and likewise for a set of
variables $\vec{X} \gets \vec{x}$. This constructs a new
\emph{counterfactual} causal model (over the same set of variables,
and consistent with the same causal graph) where the equation for each variable
$X \in \vec{X}$ is just the constant function  $X = x$.
We denote the counterfactual model constructed by
starting with $\m$ and performing intervention $\vec{X} \gets
    \vec{x}$ as $\m_{\vec{X} \gets \vec{x}}$. 


\subsection{Beliefs and Updating}

There are two sources of uncertainty: about the true model (that is,
about the causal relationship between variables) and about
the context (that is, about the realizations of
exogenous variables).  
We assume that there exists a true (but unknown) model $\m^\star$ that
is fixed across time, so that the true causal relationships between
variables remains the same at every period. By contrast, we assume that
the context is stochastic, drawn i.i.d.~at every period according to a
probability distribution over the set of contexts, $\pi \in
\Delta(\C)$. Therefore, the realization of the variables (without any
intervention) is the outcome of $(\m^\star, \c)$ where $\c$ is
distributed according to $\pi$. 
The asymmetry in how we model these two sources
of uncertainty
captures the difference between uncertainty that arises because of
persistent randomness in the system (governed by $\pi$) and the
uncertainty that arises because of the agent's epistemic state (governed by the agents beliefs about the true model, as described below). 

The agent may not know the true causal model $\m^\star$, but we assume
that the agent does know $\pi$, at least to the extent of determining those
exogenous variables she is aware of.\footnote{The set of causal models
considered by the agent may exclude some variables, including some of
the exogenous variables.  Formally, we assume that the agent knows the
marginal of $\pi$ on those variables that are included in the models
she considers possible.} Thus, if the agent learned the
true model, she would learn the true \emph{distribution} over variable
values. She would still be unable, however, to deterministically
predict the outcome, owing to the randomness of the context.%
\footnote{Conceptually, it is straightforward to also allow the agent
to be uncertain about the distribution $\pi$ over the realization of
contexts. We do not include this additional layer of uncertainty as it
increases complexity and is largely unrelated to the main focus of the
paper.} 


We now describe how we model beliefs about the true model. 
Let $E \subseteq \M$ be a non-empty subset of models. Then a
\emph{conditional probability system (cps) over $E$} is a family of
probability measures on the non-empty subsets of $E$, indexed by
non-empty subsets of $E$; that is, $\mu =
\{\mu_F\}_{\emptyset \neq F \subseteq E}$,  such that for all 
$F'' \subseteq F' \subseteq F \subseteq E$, we have
\begin{itemize}[leftmargin=15mm]
  \item[CP1.]\label{cp1} $\mu_F(F) = 1$
  \item[CP2.] $\mu_F(F'') = \mu_{F'}(F'') \times \mu_F(F')$
\end{itemize}
The measure $\mu_F$ represents the conditional beliefs given evidence of
the event $F$. The two consistency requirements state that beliefs are
as dynamically coherent as possible: (CP1) states that the evidence is treated as fact
while (CP2) extends a standard probabilistic identity to ensure
coherence for different distributions $\mu_F$.

A cps also allows beliefs conditional on events of unconditional
probability 0, where the unconditional probability is given by $\mu_E$.
To simplify notation, we denote $\mu_E$ by $\bar\mu$.
The probability $\bar{\mu}$ completely
governs the (contemporaneous) behavior of the agent, as she does not
anticipate observing inexplicable evidence; the other components of
the cps determine only her future behavior should her
contemporaneous beliefs prove incorrect.%

Shortly, we will endow the agent with the ability to take actions that
consist of intervening on some set of variables and observing the
effect of this intervention on (a different) set of variables. The
information generated by such an action is an \emph{observation};
formally an \emph{observation} is a pair $o = \<\vec{X} \gets
 \vec{x},\vec{Y} = \vec{y}\>$ where $\vec{X} \subseteq \V$ is the
set of variables that are intervened on to be set to $\vec{x} \in
\Rn(\vec{X})$ and  $\vec{Y} \subseteq \U \cup \V$ is the set of
variables that are observed to take values $\vec{y} \in
\Rn(\vec{Y})$ as a result of the intervention. We require that if
$Y\in \vec{X} \cap \vec{Y}$, 
then $\vec{y}|_Y = \vec{x}|_Y$.  Let $\O$ denote the set of all
observations. 

Let the agent's beliefs be given by a cps $\mu$.
How should the agent
update her beliefs subsequent to the observation $o = \<\vec{X} \gets
\vec{x},\vec{Y} = \vec{y}\>$? In an abuse of notation, let  
$$
\pi(\m,o) := \pi(\{\c \mid (\m_{\vec{X} \gets \vec{x}},\c) \models
\vec{Y} = \vec{y} \}) 
$$
denote the $\pi$-probability that model $\m$ would generate the
observation $o$. The agent can rule out any model not in
the set 
$$E(o) := \{\m \in \M \mid \pi(\m,o) > 0\}.$$
If the agent observes $o$, then she has \emph{learned} the event
$E(o)$.
Indeed, the agent has 
learned more: the relative likelihood of different models producing
the observation yields further belief updating. For any $F \subseteq
E(o)$, let $\mu^o_F$ be the distribution defined by 
$$\mu^o_F(\m) := \frac{\mu_F(\m)\pi(\m,o)}{\sum_{\m' \in F} \mu_F(\m')
  \pi(\m',o)},$$ 
which captures the posterior probability of models using each conditional distribution as a potential prior.

We can thus define an '`updated'' cps as
$$\mu^o := \{\mu^o_F\}_{F \subseteq E(o)},$$
which can easily be checked to be a well-defined cps over $E(o)$ whenever $E(o) \cap E \neq \emptyset$.


We conclude this section with a few special cases:

\begin{itemize}
  \item The distribution $\pi$ is trivial, that is, a point mass on a
    single context: We call such a state of affairs
    \emph{deterministic}.

  \item The supports of each $\mu_F$ is a singleton. That is, after
    any conditioning event, the agent is certain of a particular model
    and acts accordingly. 
    The results of Bramley et al.~\shortcite{bramley2017formalizing}
    suggest that this is what people actually do.
  \item The cps is over $\M$ and $\mu_\M$ is fully supported. That is,
    the agent entertains all possible models and is never surprised in
    the sense of conditioning on a 0-probability event. We could model
    such an agent without the cps machinery. 
  \item The support of each $\mu_F$ is the set of all models defined
        on a causal graph $\g$. That is, after an observation the
    agent is certain about the direction of causal influence, but not
    the exact details of the structural equations. She updates her
    beliefs about the  equations in a standard way, and only
    re-evaluates her world view when she cannot explain the data with
    her given graph $\g$. 
\end{itemize}

\subsection{Agency and Utility}\label{sec:agency+utility}

Each period, the agent can take an action; formally, an \emph{action}
is an intervention paired with a set of variables to observe: $a =
(\vec{X} \gets \vec{x},\vec{Y})$. The interpretation is that by
taking the action $a$ the agent intervenes so as to set the variables
$\vec{X} \subseteq \V$ to the values  $\vec{x} \in \Rn(\vec{X})$. In
addition, the action $a$ allows the agent to (partially) observe the
effect of the intervention; she observes the value $\vec{y} \in
\Rn(\vec{Y})$ that results from the intervention. Let $A$ denote the
set of all actions available to the agent.  

Given a setting $(\m,\c)$, each action $a =
(\vec{X} \gets \vec{x}, \vec{Y})$ determines the observation that
would be made: if $a$ is taken, then given $(\m, \c)$, the agent
will expect to observe 
$$o^{a}_{\m,\c} := (\vec{X} \gets \vec{x}, \vec{Y} = \vec{y},)$$
where $\vec{y} \in \Rn(\vec{Y})$ is the unique vector of values such
that $(\m_{\vec{X} \gets \vec{x}},\c) \models \vec{Y} = \vec{y}$. 
To avoid overly dense notation, let $\mu^a_{\m,\c} := \mu^{o^a_{\m,\c}}$ denote the c.p.s that results from updating on this observation.

Finally, let  $\util \in \V$ denote a distinguished variable which we
interpret as utility. In an abuse of notation, we can treat $\util$ as a
utility function over outcomes of variables as given by
settings. This is defined by
$\util(\m,\c) = x$ for 
the $x \in \Rn(\util)$ such that $(\m,\c) \models \util = x$. While it is
not required by what follows, it is perhaps reasonable to assume that
for all $(\vec{X} \gets \vec{x}, \vec{Y}) \in A$, we have 
$\util \in \vec{Y}$; 
that is, independent of whatever else is observed, the agent
can determine her own utility.

For an action $a = (\vec{X} \gets \vec{x}, \vec{Y}) \in A$, let $\m^a
= \m_{\vec{X} \gets \vec{x}}$, be the counterfactual model induced
by the action $a$. With this notation, we can further extend $\util$ to
capture the agent's
(one-shot) value of  taking action $a$, given her beliefs $\mu$: 
\begin{equation*}
\util(a, \mu) := \sum_{\M \times \C}  \util(\m^a,\c) \ \pi(\c) \ \bar{\mu}(\m).
\end{equation*}

The decision maker, however, does not only care about her
one-shot value but also her ability to continue to make
choices. Given a discount factor of $\delta \in (0,1)$, the agent
values the action $a$ according to its one-shot utility and its
information value. 
\begin{align*}
& \quad \val(a, \mu) := \\
&\sum_{\M \times \C} \Big( \util(\m^a, \c) + \delta  \max_{a \in A}  \val(a, \mu^a_{\m,\c})\Big)  \ \pi(\c) \ \bar{\mu}(\m).
\end{align*}
That is, the value of taking action $a \in A$ today is the expectation of
\begin{itemize}
    \item $\util(a,\mu)$ --- the immediate payoff associated with the
            intervention according to today's utility, and 
    \item $\delta \max_{a \in A}  \val(a, \mu^a_{\m,\c})$, the discounted
      value of the same problem tomorrow, except that beliefs will be
      updated according to the additional observation made.
      Recall that $o^a_{\m,\c}$ is the observation that is 
      realized when action $a$ is taken given model $\m$ and context
      $\c$, so $\mu^a_{\m,\c}$ is the agent's updated beliefs
      conditional on what she observes as part of her chosen action. 
\end{itemize}

Note that we are assuming, by nature of the ``max'' operator in the
continuation value (i.e., tomorrow's utility), that the agent
optimally uses whatever information 
arrives to maximize
the
continuation value. Note, the \emph{actual}
observation made by the agent is determined by $\m^{\star}$. So, if
$\m^\star$ is not in the support of $\bar{\mu}$, it is possible that
the true observation $o^a_{\m^\star,\c}$ does not coincide with any
$o^a_{\m,\c}$ considered with positive probability; that is,
$\bar{\mu}(E(o^a_{\m^\star,\c})) = 0$. This is \emph{not} considered
possible by the agent (i.e., it is a 0-probability event, and
therefore does not contribute to the choice of $a \in A$), but is
possible from the modeler's perspective. 
However,  the agent can still condition on the event using her cps.

\section{Steady State}

We now introduce the idea of a steady state, where the optimal action
today does not alter beliefs, so that it remains optimal
tomorrow. Therefore no new information enters the system and actions
are repeated.  
Using the notation of the previous section,
the environment (given a set of variables and ranges) is described by
the tuple
$$
(\m^\star, \pi, A, \util, \delta).
$$
Given this, a \emph{steady state} is a pair $(\mu,a)$ consisting  of a cps $\mu$, and an action $a$ such that:

\begin{itemize}
    \item $a \in \argmax_{a' \in A} \val(a', \mu)$;
    \item $\bar{\mu} = \bar{\mu}^a_{\m^\star,\c}$ for $\c \in \C$ with $\pi(\c) > 0$.
\end{itemize}

The first condition states that the agent is choosing an optimal
action \emph{given her beliefs}. The second condition states that,
given this action and the true model, the realized observation does
not change her \emph{unconditional} beliefs---hence the same action
will remain optimal in the future.  

The character of a steady state is quite straightforward in
deterministic environments (i.e., when $\pi$ is trivial), as shown
by the following two remarks, neither of which holds in the 
more general stochastic environment, as we show below. 

\begin{remark}
Suppose that $\pi$ is trivial. If $(\mu,a)$ is a steady state, then $a \in \argmax_{a' \in A} \util(a', \mu)$.
\end{remark}

\begin{proof}
Let $\c$ be the unique context in the support of $\pi$. Then, since observing $o^a_{\m^{\star}, \c}$ does not change her beliefs, it must be that $o^a_{\m, \c} = o^a_{\m^{\star}, \c}$ for every model $\m$ in the support of $\bar{\mu}$. As such, with $\bar{\mu} \times \pi$-probability 1, $\val(\,\cdot, \mu^a_{\m^\star,\c}) = \val(\,\cdot, \mu)$. The remark follows immediately. 
\end{proof}

This remark does not hold if the context
is stochastic. In particular, it is possible that an agent would
choose a statically suboptimal action because she thinks there is some
probability it will be informative (i.e., there are some contexts in
which it would allow her to discriminate between models), while in
fact (i.e., according to $\m^\star$) it is not informative for any
context.  

Our second remark shows that, in a deterministic environment,  the
agent's dynamic behavior will converge to a steady state.

\begin{remark}
\label{rmk:converge}
Assume that $\pi$ is trivial and let $\c$ be the unique context in the
support of $\pi$. 
Let $\mu$ be any cps over $E$ for some $E$ containing the true model $\m^{\star}$. 
Starting with $\mu_0 = \mu$, define inductively the
following
sequence of
actions and beliefs:  for each $i \geq 1$, take\footnote{We must
assume ties are broken consistently:  if $\val(\,\cdot\,, \mu_j)
=\val(\,\cdot\,, \mu_{j'})$ then $a_{j+1} = a_{j'+1}$.} 
\begin{itemize}
     \item $a_i \in  \argmax_{a' \in A} \val(a', \mu_{i-1})$;
     \item $\mu_i = (\mu_{i-1})_{\m^\star,\c}^{a_i}$.
 \end{itemize}
 Then $(\mu_i,a_i)$ converges to a steady state as $i \to \infty$.
\end{remark}

\begin{proof}
Given the triviality of $\pi$, $\bar{\mu}_{i} = \bar{\mu}_{i-1}$ if and only if $o^{a_i}_{\m,\c} = o^{a_i}_{\m^\star,\c} $ for every model in the support of $\bar{\mu}_{i-1}$; as such, any belief change corresponds to the agent excluding some previously considered model. Since there is a finite number of models, it follows that beliefs are eventually constant; let $k > 0$ be such that for all $i \geq k$, $\bar{\mu}_{i} = \bar{\mu}_{k}$. Then by construction we have
\begin{itemize}
    \item $a_{k+1} \in \argmax_{a' \in A} \val(a', \mu_{k})$ and
    \item $\bar{\mu}_{k} = \bar{\mu}_{k+1} =
      (\bar{\mu}_{k})_{\m^\star,\c}^{a_{k+1}}$, 
\end{itemize}
and so, a steady state.
\end{proof}

As mentioned above, Remark \ref{rmk:converge} does not hold in
general, when there is uncertainty about the context. In fact, not
only might repeated maximization not lead to a steady state, it does
not necessarily even lead to stable behavior. In other words, it is
possible that the agent's optimal action cycles between alternatives
forever. We provide a simple example of such an environment in Example
\ref{ex:no-converge}.

\section{Examples}

To get a sense of what a steady state entails, we first formalize the
examples from the introduction within this framework. We also provide
an example of the failure of convergence for non-deterministic
environments. As it is relatively simpler, we begin with Example
2.%
\footnote{In these examples, for ease of exposition, we omit the
exogenous variables, which determine the relevant endogenous variables
in the obvious way.}


\setcounter{example}{1}
\begin{example}(continued) 
There are three endogenous variables, $\V = \{\util, {\pray}, {\sun}\}$ and no exogenous variables $\U = \emptyset$.  Both (non-utility) variables have range $\{0,1\}$.  The first, 
${\pray}$, represents whether the Aitions make their daily votive
(${\pray}=1$ if they do) and the second, ${\sun}$, whether the sun
rises (${\sun}=1$ if it does). 
Consider the following two structural models (which differ only in the equation that governs ${\sun}$):
\begin{align*}
\m^\star \ &: \
 {\pray} = 0,
\quad {\sun} = 1, \\
&\quad  \util = (10\times{\sun}) - {\pray}
\end{align*}
and
\begin{align*}
\m^\dag\ &: \
 {\pray} = 0,
\quad {\sun} = {\pray}, \\
&\quad \util = (10 \times {\sun}) - {\pray}.
\end{align*}

The true model is $\m^\star$. The Aitions, however, consider $\m^\dag$. Let $\mu$ be a c.p.s such that $\bar{\mu}(\m^\dag) = 1$ and, for any $E$ with $\m^\star \in E$ and $\m^\dag \notin E$, $\mu_E(\m^\star) = 1$.

The Aitions have two actions at their disposal:
$$
a_{0} = ({\pray} \gets 0, \V) \ \ \text{ and } \ \ a_{1}
= ({\pray} \gets 1, \V).  
$$
The associated model-dependent predicted observations are:\footnote{Since there
are no exogenous variables, we suppress the subscript for the
context. Further, we abuse notation and write $o^i$ rather than
$o^{a_i}$ for the action superscript.} 
\begin{align*}
     o^{0}_{m^\star} = \<{\pray} \gets 0, ({\pray} = 0, {\sun} = 1)\>, \\
    o^{0}_{m^\dag} = \<{\pray} \gets 0, ({\pray} = 0, {\sun} = 0)\>, \\
    o^{1}_{m^\star} = \<{\pray} \gets 1, ({\pray} = 1, {\sun} = 1)\>, \\
    o^{1}_{m^\dag} = \<{\pray} \gets 1, ({\pray} = 1, {\sun} = 1)\>.
\end{align*}
The observation generated by action $a_{1}$ does
not depend on the model, whereas the observation generated by
$a_{0}$ is model dependent. Thus, starting from an initial
belief $\bar{\mu}$, the action $a_{1}$ generates an observation
that does not update beliefs at all; that is
$\bar{\mu}^{a_1}_{\m^\star} = \bar{\mu}$. On the other hand,
observing $o^{0}_{m^\star} = \<{\pray} \gets 0, ({\pray} = 0, {\sun} = 1)\>$
    cannot be explained by $\bar{\mu}$, leading to wholesale belief
    revision to $\mu_E(o^{0}_{m^\star})$, and so, given the structure of $\mu$,
    adoption of correct beliefs. 

Finally, given $\bar{\mu}$, the immediate value of each action is
given by $\util(a, \mu) = \util(\m^{\dag,a})$, so that $\util(a_{1}, \mu) =
9$, whereas $\util(a_{0}, \mu) = 0$. So, $(\mu, a_{1})$ is a
steady state: given their initial beliefs, the Aitions will continue
to pray every morning, and consequently, never update their
beliefs.\footnote{Note also that they place probability 0 on ever
receiving any information, so the choice of action is determined
entirely by the immediate value. For reference,
$\val(a_{1}, \mu) = \frac{9}{1-\delta}$, whereas $\val(a_{0}, \mu)
= 0$} 
\end{example}

In a similar manner, we can formalize Example 1:

\setcounter{example}{0}
\begin{example}(continued) 
There are five variables, $\V = \{\util, {\snake}, {\bounty}, {\eggs}, {\hunt}\}$, where ${\snake}$ represents the snake population (with range $\{0,1,2\}$), ${\bounty}$ represents the existence of a bounty (with range $\{0,1\}$), ${\hunt}$ represents whether locals hunt snakes (with range $\{0,1\}$), and ${\eggs}$ represents whether locals breed snakes (with range $\{0,1\}$).

Consider the following two structural models. 
\begin{align*}
\m^\star \ : \quad & {\bounty} = 0, \quad {\hunt} = {\bounty},\\
&{\eggs} = {\bounty}, \\
&{\snake} = 1 - {\hunt} + (2\times{\eggs}),  \\
&\util =  - (2\times {\snake}) - ({\bounty}\times {\hunt}),
\end{align*}
and
\begin{align*}
\m^\dag\ : \quad &{\bounty} = 0, \quad {\hunt} = {\bounty},\\
&{\eggsbw}  \ \textcolor{gray}{ = 0}, \quad {\snake} = 1 - {\hunt},\\
&\util =  - (2\times {\snake}) - ({\bounty}\times {\hunt}).
\end{align*}

Again, the true model is $\m^\star$, but the Governor
considers $\m^\dag$. In $\m^\dag$, the variable ${\eggs}$ is
considered irrelevant (i.e., the agent is unaware of or inattentive to
that variable).
As in Example 2, let $\mu$ be a cps  such that
$\bar{\mu}(\m^\dag) = 1$, and, for any $E$ with $\m^\star \in E$ and
$\m^\dag \notin E$, $\mu_E(\m^\star) = 1$. The Governor
must choose between
\begin{align*}
a_{1} &= ({\bounty} \gets 1, ({\snake}, {\hunt}))  \quad \text{ and  }\\
a_{0} &= ({\bounty} \gets 0, ({\snake}, {\hunt})), 
\end{align*}
with the associated predicted observations
\begin{align*}
     o^{1}_{m^\star}  = \<{\bounty} \gets 1, ({\snake} = 2, {\hunt} = 1)\>, \\
    o^{1}_{m^\dag}  = \<{\bounty} \gets 1, ({\snake} = 0, {\hunt} = 1)\>, \\
    o^{0}_{m^\star} = \<{\bounty} \gets 0, ({\snake} = 1, {\hunt} = 0)\>, \\
    o^{0}_{m^\dag} = \<{\bounty} \gets 0, ({\snake} = 1, {\hunt} = 0)\>.
\end{align*}
As in the Example 2, the observation generated by one action (here
$a_{0}$) is model independent, whereas the observation generated
by the other action (here $a_{1}$) depends on the model. In
contrast to Example 2,
however, the action chosen under the initial beliefs is the revealing
action. To see this, note that, since the Governor places
probability 1 on $\m^\dag$, he sees no value in exploration, and thus
chooses his action based only on the immediate payoff.
Since $\util(a_{1},  \mu) = -1$ and
$\util(a_{0},  \mu) = -2$, the Governor institutes
the bounty, expecting the snake population to
decline. However, the actual observation is $o =
o^{1}_{m^\star}$: an increase in the snake population (and a
realized utility of $-5$). 

This observation cannot be explained by his current model, so the Governor would revise his beliefs to $\mu_{E(o)}$. Under his new beliefs, his optimal action is to retract the bounty, which corresponds to steady-state behavior. 
\end{example}

Notice that in this example, when the value of ${\eggs}$ is fixed to
0, the equations of $\m^\star$ reduce to the equations of
$\m^\dag$. That is, one could view $\m^\dag$ as a specification of
$\m^\star$ where a particular variable is ignored (i.e., is implicitly
deemed unchangeable, is believed to bear no causal relation to any
utility-relevant variable, or lies outside of the agent's awareness).
Interpreting $\m^\dag$ as capturing the beliefs of an
agent who is unaware of ${\eggs}$ and $\m^\star$ as encoding her
understanding after becoming more aware, then this equational relation
between the two models embodies a principle akin to the principle of
\emph{Reverse Bayesianism} \cite{karni2013reverse} in state-space
models of awareness. That is, when awareness expands, the
previously-held relationships between variables remain as a special
case of the new expanded relationships.  

\setcounter{example}{2}
\begin{example}
\label{ex:no-converge}
Let $\U = \{Y\}$ be a single variable with range $\Rn(Y) = \{0, 1, \ldots, 99\}$, and $\pi$ the uniform distribution over outcomes. There are two endogenous variables, $\V = \{\util, X\}$, where $X$ is the only variable that can be intervened on, and 
$\util$ 
is the agent's utility with ranges $\Rn(X) = \{r,l,m\}$ and $\Rn(\util) = \{0,1\}$. We consider three causal models:
\begin{align*}
\m^r \ :  \quad X = m, \quad \util = \begin{cases}
1 \text{ if } X = r \text{ and } Y \geq 25 \\
1 \text{ if } X = l \text{ and } Y \geq 50 \\
0 \text{ otherwise; }
\end{cases}
\end{align*}

\begin{align*}
\m^l \ :  \quad X = m, \quad \util = \begin{cases}
1 \text{ if } X = r \text{ and } Y \geq 50 \\
1 \text{ if } X = l \text{ and } Y \geq 25 \\
0 \text{ otherwise; }
\end{cases}
\end{align*}

\begin{align*}
\m^\star \ :  \quad X = m, \quad \util = \begin{cases}
1 \text{ if } Y \geq 50 \\
0 \text{ otherwise. }
\end{cases}
\end{align*}
Suppose that there are two actions:
$$a_r = (X\gets r, U) \text{ and  } a_l = (X\gets l, U).$$

Note the following simple observations: first, for beliefs
sufficiently concentrated on $\m^r$ (resp., $\m^l$), the action $a_r$
(resp., $a_l$) maximizes expected utility; second, since the agent cannot
observe $Y$, there are no inexplicable outcomes; all three models can
produce any combination of $X$ and $U$. As such, she will never
completely discard any model that is in the support of her initial
beliefs. 

Finally, assume the support of the agent's initial beliefs,
$\bar{\mu}$, is $\{m^r,m^l\}$.  In this case, her behavior
will switch between the two actions infinitely often. To see this,
assume to the contrary that her behavior was eventually
constant. Without loss of generality, assume she eventually chooses
$a_r$ forever. Under the true model $\m^\star$, this produces the
outcome $(X \gets r, \util = 1)$ with probability $\frac12$ and the
outcome $(X \gets r, \util = 0)$ with probability $\frac12$.
This is the expected distribution of observations under $\m^l$,
but \emph{not} under $\m^r$; thus, with $\pi$-probability 1, with
sufficiently many observations, her beliefs will become arbitrarily
concentrated on $\m^l$; however, for such beliefs, $a_r$ is no longer
optimal. 
\end{example}

\section{Introspective Unawareness}

The agents in the framework laid out above entertain a classical
exploration/exploitation  trade-off: given their initial uncertainty
about the true causal model, the forward looking agents maximizing
$\val(\,\cdot, \mu)$ can take statically suboptimal actions to gather
information about the causal structure, enabling better future
decisions. The perceived value of exploration is determined by the
agent's current beliefs, $\bar{\mu}$, and so she acts with certainty
that the true model is contained in the support of $\bar{\mu}$. This
is true even if the agent has repeatedly abandoned prior beliefs in
the face of surprising observations; she never considers the
possibility that her \emph{current} beliefs might be misspecified or
incomplete. 

In this section, we augment the model above slightly to accommodate an
introspectively unaware agent, one who entertains the possibility that
her beliefs are incomplete. Of course, if the agent was aware of which
relevant models she was failing to consider, she could simply shift
some small probability mass onto these models, expanding the support
of her beliefs to capture all relevant possibilities. By way of contrast, we
assume that the agent is (a) unaware of the models outside the support
of $\bar{\mu}$, but (b) aware that she is unaware. 
This parallels awareness of unawareness in state-space models \cite{karni2017awareunawareness}.
The agent
therefore cannot predict what it is that she might observe that would
make her abandon her current world view.  

In essence, the idea is that the agent considers the possibility of encountering some inexplicable evidence, and conjectures about the potential likelihood and value of doing so, but cannot explicitly determine what the world would look like afterwards. 
To model this, we introduce two new objects into our model. 

\begin{itemize}
 \item First, let $v^*$ denote the agent's subjective value of
   learning something unforeseeable. This represents the agent's best
   guess about the (discounted expected) value of a paradigm shift
   that causes her to abandon her current beliefs.  
    \item Let $\xi: A \to [0,1]$ denote a function that represents the
    agent's beliefs about the likelihood of each action yielding a
    surprising outcome, and let $\tau: [0,1]^A \times \O \to
        [0,1]^A$ describe how the agent
    updates her beliefs about the likelihood of surprise.  
  
\end{itemize}

With these two pieces of notation, we can consider the introspective agent's version of $\val$:

\begin{align*}
&\val^*(a, \mu, \xi) = \xi(a) v^* + (1\,{-}\,\xi(a)) \sum_{\M \times \C} \Big( \util(\m^a, \c)  \\
&+ \delta  \max_{a \in A}  \val^*(a, \mu^a_{\m,\c}, \tau(\xi, o^a_{\m^\star,\c}))\Big)  \ \pi(\c) \ \bar{\mu}(\m).
\end{align*}
That is, the value of an action $a$ is the $\xi(a)$ convex combination
of $v^*$---the value subjectively assigned to unknown outcomes---and
the expected value of the problem under the distribution
$\bar{\mu}$. This second component is \emph{almost} the same as $\val$ as
defined in Section~\ref{sec:agency+utility}, except that the
continuation value also 
allows for the possibility of discovering a novel outcome, where the
probabilities of discovery have been updated by $\mu$. 

Notice that the appeal of the novel outcome only kicks in when the known part of the state-space provides low expected payoff. 
Therefore, in line with the finding of Bonawitz
et al.~\shortcite{bonawitz2014win}  
the actions that yield low ``objective'' payoff but high chance of revealing something novel will only be desirable when actions with higher objective payoff are unavailable. 

We can therefore recast our notion of a steady state using $v^*$: Call
a tuple $(\mu,a, \xi)$ an \emph{introspective steady state} if
\begin{itemize}
    \item $a \in \argmax_{a' \in A} \val^*(a', \mu, \xi)$
    \item  $\bar{\mu} = \bar{\mu}^a_{\m^\star,\c}$ for $\c \in \C$ with $\pi(\c) > 0$, and
    \item $\tau(\xi, o^{a}_{m^\star, \c}) = \xi$ for $\c \in \C$ with $\pi(\c) > 0$.
\end{itemize}
The first two conditions are direct analogs of the conditions of a
(non-introspective) steady state. The third also requires that the
probabilities of discovery remain unchanged.



\section{Conclusion}
We have introduced a formal model of updating causal beliefs, which
can capture the way people seem to update their beliefs in a
straightforward way.  The model allows us to consider optimal
decision-theoretic rules for updating, and explains how people can
maintain incorrect steady-state beliefs.  In future work, we plan extensions to  
game-theoretic considerations, and to examine how agents can use these
ideas strategically (e.g., to mislead other agents).

\appendix




\section*{Acknowledgments}

We would like to thank seminar audiences at the Philosophy, Politics, and Economics Society London Meeting and at Time, Uncertainties, and Strategies XI.

\clearpage

\small
\bibliographystyle{named}
\bibliography{joe}

\end{document}